# 2.5-kV AlGaN/GaN Schottky Barrier Diode on Silicon Substrate with Recessed-anode Structure

Ru Xu, Peng Chen*, Menghan Liu, Jing Zhou, Yunfei Yang, Yimeng Li, Cheng Ge, Haocheng Peng, Bin Liu, Zili Xie, Rong Zhang, Youdou Zheng

Key Laboratory of Advanced Photonic and Electronic Materials,
School of Electronic Science and Engineering, Nanjing University, Nanjing 210093, China

*Abstract*—In this letter, we demonstrate high-performance lateral AlGaN/GaN Schottky barrier diodes (SBD) on Si substrate with a recessed-anode structure. The optimized rapid etch process provides results in improving etching quality with a 0.26-nm roughness of the anode recessed surface. By using the high work function metal Pt as the Schottky electrode, a low $V_{on}$ of 0.71 V is obtained with a high uniformity of ± 0.023 V for 40 devices. Supported by the flat anode recess surface and related field plate design, the SBD device with the anode-cathode spacing of 15 μm show the $R_{on,sp}$ of 1.53 mΩ·cm² only, the breakdown voltage can reach 1592 V with a high power FOM (Figure-of-Merit) of 1656 MW/cm². For the SBD device with the anode-cathode spacing of 30 μm, the breakdown voltage can be as high as 2521 V and the power FOM is 1244 MW/cm².

## I. INTRODUCTION

GROUP III nitrides represented by GaN have a wide range of applications in optoelectronics and power electronics, which is considered to be one of the most important semiconductor materials after silicon. Owing to its superior material properties such as high electric breakdown field, high-electron saturation velocity, and high mobility in a readily available heterojunction 2-dimensional electron gas (2-DEG) channel, GaN has been widely used in high frequency and power electronic devices. Power electronic devices are one of the fundamental components in almost all electronic manufacturing industries, including the field of consumer electronics, wireless communications and industrial control. Among all device types, Schottky barrier diodes (SBD) are the essential components in power conversion systems, such as switching power supply, frequency converter, driver and other circuits.

Recently, high-performance GaN-based lateral SBD devices have received sufficient attention and many results have been reported, such as the SBDs with low $V_{on}$ of 0.2~0.8 V, the breakdown voltages from 1 kV to 2.7 kV and a high-power figure-of-merit (FOM) of 2.65 GW/cm² [1]-[22]. According to the previous reports, in order to reduce the $V_{on}$, anode recess structure has been widely used for GaN lateral SBDs [4] [5]. Obviously, a good surface condition of the anode recess is the key factor to determine the final performance of the SBD. Most of the reported results adopted a complicated etch process to ensure controllable recess depth and minimize etch damage [1] [5] [6] [16] [21]-[23]. In order to realize an effective process and a good surface condition, a rapid anode recess etching method is proposed in this work, which can not only obtain a relatively precise recess depth, but also obtain a flat etching surface. Based on this etching technique and the special anode structure optimized in this work, high performance SBD devices can be demonstrated.

## II. DEVICE FABRICATION

The sample was grown by metal organic chemical vapor deposition (MOCVD) on a 2-inch Si substrate. From the substrate, the device structure consists of a 5-μm high-resistance GaN buffer layer, a 200-nm i-GaN channel layer, a 1-nm AlN insertion layer, a 20-nm Al$_{0.25}$Ga$_{0.75}$N barrier layer, a 2-nm carbon doped GaN cap layer and a 30-nm in-situ SiN$_x$ passivation layer.

The SBD fabrication commenced with photolithography defining the SBD isolation area, and then putting the sample into buffered oxide etch (BOE) for 5 minutes to remove the SiN$_x$, finally a 400 nm isolation mesa was etched by inductively coupled plasma (ICP) by using Cl$_2$/BCl$_3$.

In the second step, the electrode recess area was first defined by photolithography, and then the sample was put into BOE for 5 minutes to remove the SiN$_x$. The power of the RF/ICP was selected as 100W/300W during the ICP etching. The duration of the etching was 10 s, resulting in the etching depth about 25~27 nm measured by an atomic force microscope (AFM). Next, the sample was etched in a diluted KOH solution for 15 minutes in a water bath at 80 °C to remove all surface damages.

The last step was evaporation of electrodes, we used Ti/Al/Ni/Au (30/150/30/100 nm) as the ohmic electrode followed by rapidly annealing at 850 °C for 30 s in N$_2$. The ohmic contact resistance ($R_C$) and sheet resistance ($R_{SH}$) were

This work is supported by National High-Tech R@D Project (2015AA033305), Jiangsu Provincial Key R&D Program (BK2015111), Collaborative Innovation Center of Solid State Lighting and Energy-saving Electronics, the Research Funds from NJU-Yangzhou Institute of Opto-electronics, the Research and Development Funds from State Grid Shandong Electric Power Company and Electric Power Research Institute. (*Corresponding authors: pchen@nju.edu.cn)

This work has ever been successfully submitted to the IEEE Electron Device Letters for possible publication in Sep. 2019, and the manuscript ID is EDL-2019-09-2015.

0.67 Ω·mm and 323 Ω/□, respectively. Next, the anode metal Pt/Au (50/300 nm) was evaporated on the recessed area, which has a 2-μm overlap on the SiN$_x$ passivation as the field plate.

Fig. 1.(a) shows the device structure. The SBDs feature a circular configuration with an effective width 220 μm. The anode-cathode spacing ($L_{AC}$) varies from 5 μm to 30 μm and the anode recess depth ($d_R$) varies from 12 nm to 33 nm. The internal electric field distribution of the SBD is shown in Fig. 1. (b). When there is no anode field plate (AFP), the electric field ($EF$) will concentrate at the edge of the anode at $V_{anode}$ = -300 V. The result is that there is only one anode peak $EF$ and the field intensity is very high at that spot. However, when an AFP is

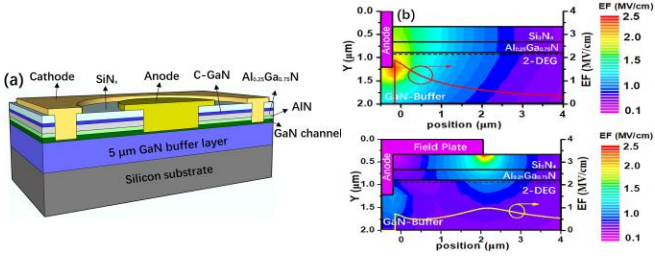

Fig. 1. (a) Cross-section view of the SBD device (b) internal electric field distribution of SBDs modulated by AFP

added to the anode, at $V_{anode}$ = -300 V, the peak $EF$ at the anode edge is reduced, the AFP dispersed the electric field, ultimately increasing the breakdown voltage of the SBD.

The SBD devices were analyzed by Silvaco-TCAD for the internal electric field distribution. The forward direct current (DC) measurements were carried out with a Keithley 4200 Semiconductor Parameter Analyzer and the reverse breakdown measurements were performed using an Agilent B1505A high-voltage semiconductor analyzer system.

## III. RESULTS AND DISCUSSIONS

For the anode recess etching process, various etching methods have been reported, including high-temperature ICP etching, digital etching (dry and wet etching oxide layer circulation), two-step etching (ICP combined with digital etching), etc [24]-[26]. The advantages of these methods are that they can control the etching depth accurately and guarantee low etching damage. Although the controllability of the etching is increased, the efficiency of these methods is extremely low. For GaN E-mode HEMT, the accuracy of gate recess depth is very important. It requires a thin AlGaN barrier for gate to deplete the 2-DEG. This means that the AlGaN layer cannot be completely etched, which requires a high controllability of etching speed. However, for AlGaN/GaN SBD, the accuracy of etching depth is not that important, as long as the anode is in direct contact with the GaN channel, and don't go deep into the GaN buffer. Reports and our experiment have shown that different etching depth of anode has no deep influence on the performance of the device [27], because it is the conduction between anode and cathode, the AlGaN below the anode is not that important. This makes the range of anode recess depth larger. At this time, using the above method will no doubt increase the complexity of the device fabrication process.

However, most of the reported high-performance AlGaN / GaN SBDs still use the above inefficient anode recess etching methods [1] [5] [6] [16] [21]-[23]. Thus, in order to develop an efficient etching method for recessed anode to improve the performance of the SBD, we did two different experiments. They were rapid etching method and slow etching method, the conditions were RF/ICP-100/300W and RF/ICP-10/100W, the etching time were 10s and 1min50s, respectively, the etching gas was a mixture of Cl$_2$ / BCl$_3$, the flow rate was 48 / 6 sccm, and the chamber pressure was 10 mtorr. Both of the etching depth are ~25nm, after etching, we put all the samples into a 0.1 mol/L KOH solution at 80 °C water bath for 15 min to remove the etch damage.

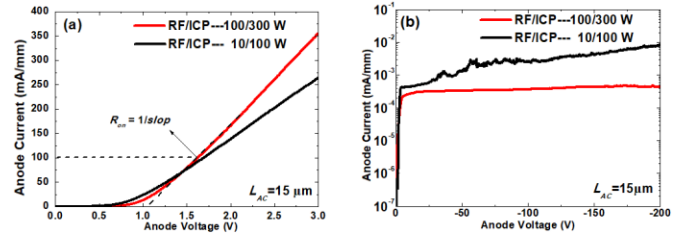

Fig. 2. I-V characteristics of two different anode-recess etching conditions (a) RF/ICP 100/300W (b) RF/ICP 10/100W (both after KOH solution wet etch)

As shown in Fig. 2, two different etching methods show different results. For the rapid etching process, the $V_{on}$ of the device is 0.71 V ($I_F$=1 mA/mm). Considering a 1.5-μm transfer length of ohmic contact and a 1.5-μm extension length of the Schottky contact, with the equation $R_{on, sp} = R_{on} × (L_{AC} + 3)$, $R_{on, sp}$ is 1.53 mΩ· cm$^2$ ($I_F$=100 mA/mm). But for the slow etching process, the $V_{on}$ of the device drops to 0.49 V, $R_{on, sp}$ rises to 2.48 mΩ· cm$^2$, and the leakage current also rises by more than an order of magnitude ($V_R$=-200 V), which means the slow-etching method has a larger damage to the surface and sidewall.

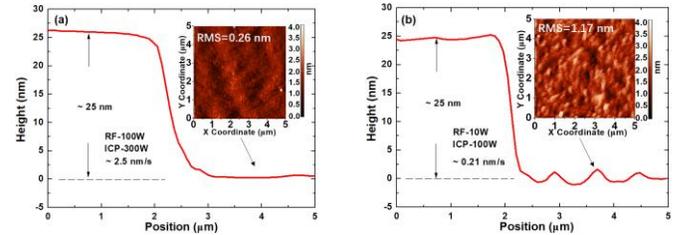

Fig. 3. AFM surface of two different anode-recess etching conditions (a) RF/ICP 100/300W (b) RF/ICP 10/100W (both after KOH solution wet etch)

Although both of them have been modified under the same KOH condition, they show different results. In order to find out the cause of this phenomenon, we tested the surface of anode recess by AFM, as shown in Fig. 3, according to known reports, the slower the etching speed, the smoother the etching surface [5]. Although the etching speed has decreased, the damage caused by etching has not been removed, to remove the damage, we added a 0.1 mol/L hot KOH solution treatment after etching. Interestingly, after the treatment of KOH solution, the surface RMS of rapid-etched sample is only 0.26 nm, but the surface RMS of slow-etched sample is 1.17 nm, which shows that a high-power etching with a short time can also obtain a flat surface.

Since hot KOH solution cannot fully remove the damage of slow etching, in order to further remove the etch damage, we put the slow-etched SBD into a rapid thermal annealing(RTA) for low-temperature annealing at a $N_2$ atmosphere, the temperature were 300 °C, 400 °C, 500 °C, respectively, and the annealing time was 5 min. The test results are shown in Fig. 4, we can find that the forward characteristics of slow-etched samples are significantly restored after RTA at low temperature compared with those of non-annealed samples, $R_{on,sp}$ recovers to the same level as the rapid-etched samples, which means the density of 2-DEG is also restored. However, for the reverse characteristics, we find that low-temperature post annealing could reduce the leakage current to a certain extent. As the annealing temperature rises from 300 °C to 500 °C, the leakage current of the SBD first decreases, and then rises to the level of the non-anneal sample. The effect of damage repair at 400 ° C is the best, this is mainly because the N vacancy on the damaged surface is repaired by low-temperature annealing in the $N_2$ atmosphere, but the further rise of temperature can introduce new N vacancy on the GaN surface [28], so it is very important to choose the appropriate post annealing temperature. However, despite we did the post annealing experiments, the leakage current of the slow-etched sample is still larger than that of the rapid-etched sample.

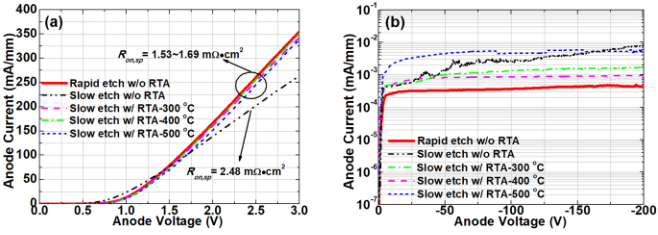

Fig. 4. *I-V* curves of different post annealing conditions (a) forward (b) reverse

Through the above experimental analysis, we selected RF/ICP-100/300W anode recess etching with a hot diluted KOH solution treatment as the condition of final device fabrication process.

As mentioned earlier in this paper, to explore the influence of different anode recess depth on the *I-V* characteristics of the SBD, four groups of devices with different anode recess depths were fabricated and analyzed, the recess depths of the four groups are non-recessed, 12 nm, 25 nm and 33 nm, respectively.

As shown in Fig.5, The non-recessed sample and the sample with the AlGaN remained ($d_R$=12 nm) draw a large $V_{on}$, both of which are more than 1V. When the etching depth reaches 25nm and 33nm, which means that the AlGaN layer is completely etched, and go into the GaN channel layer. As can be seen from the figure, when the anode contacts with the GaN channel directly, the $V_{on}$ reaches the minima. There is no difference in the $V_{on}$ between the two samples, both of which are 0.71 V. We extracted the $R_{on,sp}$ of three SBDs in the same wafer with $d_R$ = 33 nm, which ranges from 1.53 m$\Omega$·cm$^2$ to 1.86 m$\Omega$·cm$^2$ and the variation of which is within the acceptable range of error.

As for the leakage current, the largest leakage current is found in the non-recessed sample, due to the lack of post modification, the in-situ defects and surface states always exist, resulting in the larger leakage current of the device. The leakage current of recessed and post-modified samples is kept in the order of $10^{-4}$ mA/mm, which is two orders of magnitude lower than the reported 2.07 kV GaN-on-Si sample [5].

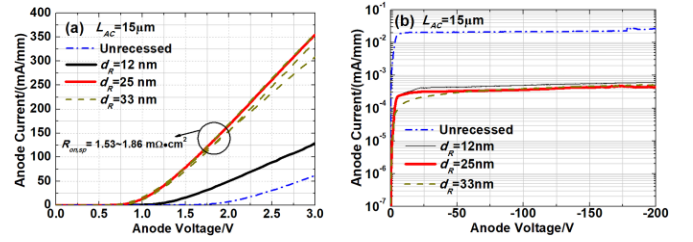

Fig. 5. Forward characteristics (a) and leakage current (b) for the devices with different anode recess depths $d_R$ at room temperature.

Fig. 6. (a) shows the forward characteristics of 40 sets of $L_{AC}$=15 μm SBDs with $d_R$ of 25 nm on two different wafers. Fig. 6. (b) is the histogram statistics of $V_{on}$, the average $V_{on}$ of the 40 SBDs is 0.71 V, while exhibiting a small standard deviation of

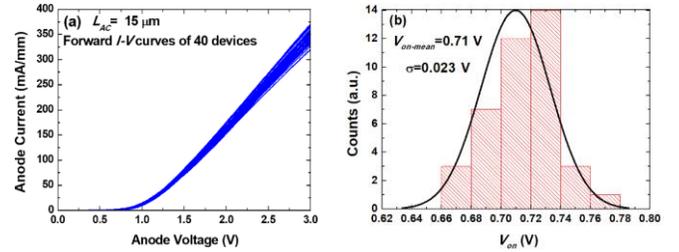

Fig. 6. *I-V* curves of 40 devices on two different wafers with $L_{AC}$= 15 μm (a) and distribution of $V_{on}$ (b).

0.023 V. The $R_{on,sp}$ ranges from 1.44 m$\Omega$·cm$^2$ to 1.79 m$\Omega$·cm$^2$ .It shows the extreme uniformity of the anode recess surface and demonstrates high repeatability of our rapid anode recess etch process.

Based on above results, considering both the forward and reverse characteristics, the SBD devices with anode recess depth of 25 nm have the best performance, which will be used in following study.

Fig.7 shows the forward *I-V* characteristics of the SBDs with different $L_{AC}$ values (from 5 μm to 30 μm). The normalized forward operating current of the SBD with $L_{AC}$ = 5 μm and 15 μm can reach 237 mA/mm and 168 mA/mm at a forward bias of 3 V, respectively. Combined with a low leakage current below 0.0003 mA/mm, the current on/off ratio of these devices can be more than $10^8$ within the range of − 3 V to 3 V on the anode at room temperature. At the forward current density of 100 mA/mm, the $R_{on,sp}$ values of the SBDs with $L_{AC}$=5 μm and $L_{AC}$=15 μm are 0.45 mΩ·cm² and 1.53 mΩ·cm², respectively.

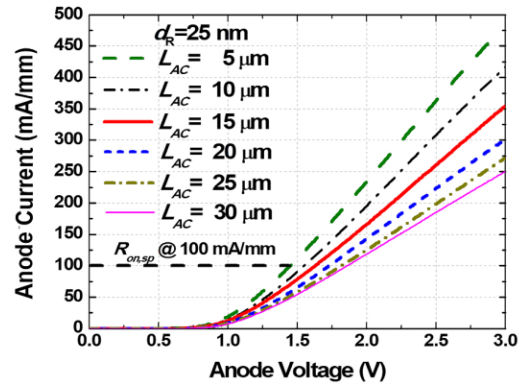

Fig. 7. Forward *I-V* characteristics of the SBDs with different $L_{AC}$ of 5 μm, 10 μm, 15 μm, 20 μm, 25 μm and 30 μm.

For Schottky forward characteristics, the I-V characteristic can be expressed by the theory of thermionic field emission (TFE) as:

$$I = I_s \times \left[ \exp\left( \frac{qV - I \times R_s}{nkT} \right) - 1 \right] \quad (1)$$

Where $R_s$ is the series resistance, $T$ is the absolute temperature, $k$ is the Boltzmann constant, $n$ is the ideal factor, and $I_s$ is the reverse saturation current. $n$ and $I_s$ can be obtained by the slope and intercept of the ln $I$-$V$ curve, respectively. Meanwhile, the Schottky barrier height $\varphi_b$ can be expressed as:

$$\varphi_b = \left( \frac{kT}{q} \right) \times \ln\left( \frac{AA^*T^2}{I_s} \right) \quad (2)$$

Where $A$ is the effective area of the device and $A^*$ is the Richardson constant, after calculation, $n$ and $\varphi_b$ are 1.2 and 0.58 ± 0.04 eV, respectively.

The dynamic performance of the SBD is shown in Fig.8(a), the anode reverse pulse was set to 0~-800V with a step of -200V, pulse time was 30 ms, then, anode voltage was switched to positive values from 0~3 V with a step of 0.05 V to turn on the SBD. Fig. 8(b) shows the dynamic $R_{on,sp}$ of the SBDs with 15 μm $L_{AC}$ under quiescent reverse bias from 0~-800 V with a stress time of 30 ms. there is only 0.12 V $V_{on}$ shift when compared with the one without reverse stress under the same

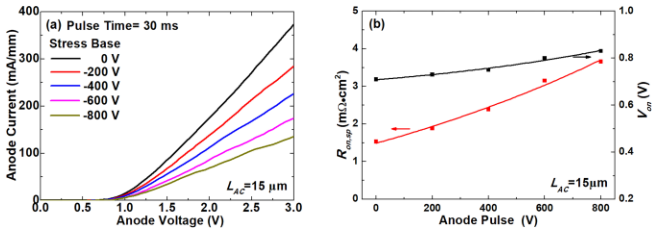

Fig. 8. (a) forward I–V characteristics of the GaN-on-Si SBDs with 30 ms time under various stress bases from -200 to -800 V at $L_{AC}$=15 μm. (b) Extracted dynamic $R_{on,sp}$ and $V_{on}$ versus the reverse stress voltages

measurement setup. The dynamic $R_{on,sp}$ at -800 V reserve stress voltage is about 2.4 times the one without reverse pulse, which indicates that further optimization of passivation layer are needed.

Fig. 9.(a) shows the reverse I-V characteristics of the SBD at room temperature, with the $L_{AC}$ changing from 5 μm to 30 μm. When a destructive-breakdown occurs, the voltage applied to the anode is defined as the breakdown voltage. The results show that the breakdown voltages of our SBD devices are 615 V ($L_{AC}$=5 μm), 1116 V ($L_{AC}$=10 μm), 1592 V ($L_{AC}$=15 μm), 1837 V ($L_{AC}$=20 μm), 2306 V ($L_{AC}$=25 μm) and 2521 V ($L_{AC}$=30 μm) respectively. We also tested five sets of the SBDs

with $L_{AC}$=25 μm, the breakdown voltages ($BV$) are all around 2300 V. Considering a 1.5-μm transfer length of ohmic contact and a 1.5-μm extension length of the Schottky contact, the power FOMs (FOMs = $BV^2/R_{on,sp}$) are 840, 1354, 1656, 1303, 1445 and 1244 MW/cm² for the SBDs with $L_{AC}$= 5, 10, 15, 20, 25 and 30 μm respectively. Our lateral GaN SBDs are benchmarked against some state-of-the-art lateral GaN SBDs (as shown in Fig. 9.(b)) [1]-[22].

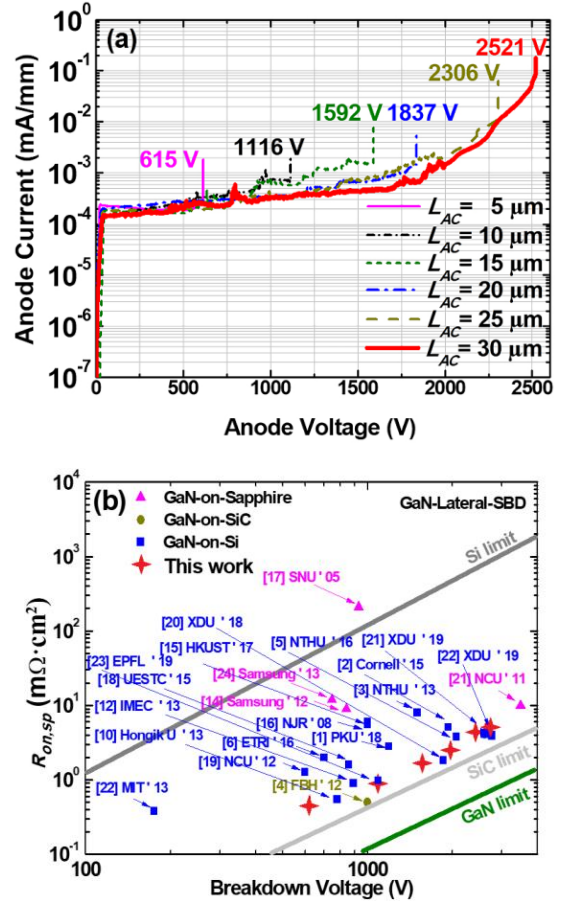

Fig. 9. (a) Reverse blocking characteristics of the SBDs as a function of $L_{AC}$ from 5 μm to 30 μm and (b) benchmark plot of $BV$ versus $R_{on,sp}$ for our lateral GaN-based SBDs.

## IV. CONCLUSIONS

In this work, we developed the high-performance lateral SBDs by introducing a rapid anode recess etch technology to replace the conventional slow anode recess etch technology. The obtained anode recess surface has an RMS of only 0.26 nm and shows great improvement on the device performance. The repeatability test results show that the $V_{on}$ is 0.71 V ± 0.03 V among 40 SBD devices. When $L_{AC}$=15 μm, the breakdown voltage can reach 1592 V, $R_{on,sp}$ is 1.53 mΩ·cm², and the corresponding power FOM is as high as 1656 MW/cm². By using this approach, the highest breakdown voltage is up to 2521 V, obtained in the SBDs with $L_{AC}$=30 μm, and the power FOM is 1244 MW/cm² correspondingly. Our results indicate that it is possible to achieve high-repeatability GaN SBDs with higher breakdown voltage on Si.